# Slow Molecules Produced by Photodissociation

Bum Suk Zhao, So Eun Shin, Sung Tai Park, Xingnan Sun, and Doo Soo Chung[*]

*Department of Chemistry, Seoul National University, Seoul 151-747, Korea*

**Abstract**

A simple method to control molecular translation with a chemical reaction is demonstrated. Slow NO molecules have been produced by partially canceling the molecular beam velocity of $NO_2$ with the recoil velocity of the NO photofragment. The $NO_2$ molecules were photodissociated using a UV laser pulse polarized parallel to the molecular beam. The spatial profiles of NO molecules showed two peaks corresponding to decelerated and accelerated molecules, in agreement with theoretical prediction. A significant portion of the decelerated NO molecules stayed around the initial dissociation positions even several hundred nanoseconds after their production.

**I. Introduction**

During the last three decades, atomic physics has developed means to control atomic translational motion. Atoms have been deflected, focused, decelerated, cooled, trapped, and Bose-Einstein condensed [1], and a prototype atom laser has been implemented [2]. There has also been a considerable effort made in molecular physics to gain control over molecular translation, as molecules offer a wide range of properties not available with atoms. However, laser cooling, the workhorse of cold-atom physics, is not applicable to molecules. This is because the requisite closed absorption-emission cycle is essentially impossible to realize in molecules, due to their complicated internal energy-level structures.

Currently, a number of direct and indirect techniques have been developed for the slowing and cooling of molecules based on a variety of principles. The indirect techniques make use of translationally cold atoms produced by laser cooling and bring them together either via photoassociation [3] or collisions controlled via Feshbach resonances [4, 5]. Direct techniques slow down or cool preexisting molecules; this is done by buffer-gas cooling [6], Stark deceleration [7], counter-rotating-nozzle slowing [8], single collision scattering [9], or light-field slowing [10].

In addition, the motion control of molecules should also deal with rotational degrees of freedom since the molecular anisotropy gives rise to a strong dependence of molecular interactions and properties on their spatial alignment and orientation. A supersonic expansion can be used to align the rotational angular momentum perpendicular to the flight direction [11, 12]. An electric hexapole field [13] or a strong DC field [14] can be used to produce states in which the molecular axis is oriented with respect to the electrostatic field. Also, an intense laser field can be used to align the molecular axis along the laser's polarization vector [15].

One of the main motivations for controlling molecular translation and rotation comes from chemical reaction dynamics [15]. While molecular orientation/alignment is crucial for elucidating steric effects, cold/slow molecules can open up the study of cold or ultra-cold chemistry. At extremely low translational energies, the de Broglie wavelength may exceed molecular dimensions. This may entirely change the nature of reaction dynamics. At such low temperatures, although far from reality yet, even collisions of large molecules exhibit significant quantum effects.

In this work, a photodissociation reaction is used as a relatively simple method to produce slow molecules. It takes advantage of the recoil energy released into the translation of molecular photofragments. This energy can be either subtracted (deceleration) or added (acceleration) to the velocity of the photofragments, depending on the orientation of the molecular axis with respect to the molecular beam axis and the photodissociation laser polarization vector. This mechanism is well known in reaction dynamics and expected to produce zero velocity molecules when the proper condition of kinematics is fulfilled. Therefore, this method is even simpler than the relatively uncomplicated techniques such as the single collision scattering [9] and the light-field slowing [10]. In a proof-of-principle experiment, slow NO molecules were produced from the photofragmentation of $NO_2$. Since the angular distributions of the photofragments are quite anisotropic, a significant yield of decelerated or accelerated photofragments could be obtained. This anisotropy can be enhanced with a help of the various alignment technique mentioned above [11-15].

## 2. Experimental

The experimental setup of the previous molecule optics experiments [16-18] was used without major modifications (Fig. 1a). The vacuum system consisted of two differentially pumped chambers. A neutral supersonic molecular beam of 240 $\mu$s duration was formed by expanding 79 torr $NO_2$ mixed with argon to a total pressure of 2 atm into the source chamber at a rate of 10 Hz. The molecular beam passed through a 0.51-mm-diam skimmer to the detection chamber and then through a 0.6-mm-diam pinhole. The linearly polarized third harmonics of a Nd:YAG laser pulse of 7.1 ns duration and 29 $\mu$m waist radius was focused into the pulsed molecular beam and dissociated $NO_2$ into NO and O through the following one photon reaction: $NO_2 + h\nu(355 \text{ nm}) \rightarrow NO(X^2\Pi, v = 1) + O(^3P_J)$. The laser focus center, laser propagation direction, and molecular beam axis were chosen as the origin, $x$ axis, and $z$ axis, respectively. The intensity and the laser polarization direction of the 355 nm UV laser were controlled by the combination of a Glan-laser polarizer and a half wave plate. The NO molecules were then ionized by a dye laser pulse through a resonance enhanced multiphoton-ionization (REMPI) process. The dye laser was pumped by the third harmonics of another Nd:YAG laser and adjusted to generate pulses of 473 nm wavelength, 5 ns pulse width, and 0.3 mJ/pulse energy. At 473 nm, NO in the first vibrationally excited state was ionized via a [2+2] REMPI process based on the [1+1] REMPI process at ~236 nm [19, 20]. The two laser beams were focused by a lens of 20-cm focal length and their relative positions adjusted by tilting a mirror for the UV laser with a differential micrometer to avoid time-of-flight (TOF) variations from the changes in the probe dye laser positions. The NO$^+$ ions were then accelerated and focused by three gridless electrodes, a repeller, an extractor, and a ground, onto a microchannel plate (MCP) after flying 67 cm through a TOF tube. These electrodes perform velocity map imaging [21]. Ion signals amplified by the MCP hit a phosphor screen at its back to result in luminescence, which was detected by a photomultiplier tube and a gated intensified charge coupled device camera.

## 3. Results and discussion

When $NO_2$ molecules are dissociated into NO and O by a 355 nm laser pulse, the dissociation probability is linearly proportional to the laser intensity, since it is a one-photon process. With a Gaussian profile laser pulse of waist radius $\omega_0$ and pulse width $\tau$ propagating along the $x$ axis, the dissociation probability at position $A_0(x_0, y_0, z_0)$ and time $t_0$, $p(x_0, y_0, z_0, t_0)$, is proportional to $(1/\tau\omega_0^2)\exp[-(2y_0^2 + 2z_0^2)/\omega_0^2]\exp[-(4\ln2)t_0^2/\tau^2]$. The time $t = 0$ is defined as the moment of maximum intensity of the dissociation laser at the origin $O(0,0,0)$ (Fig. 1b). If the laser is linearly polarized, the photofragments will have an angular distribution in terms of the angle $\theta$ between the laser polarization and the recoil velocity [14, 22],

$$I(\theta) = \frac{1 + \beta P_2(\cos\theta)}{4\pi}, \qquad (1)$$

where $\beta$ is the recoil anisotropy factor and $P_2(x) = (3x^2 - 1)/2$. Then the number of NO molecules $N(z, t; x_0, y_0, z_0, t_0, v_{rec})$, which are produced at $A_0$ and $t_0$ with a recoil speed $v_{rec}$ and arriving at the probing position $P(0, 0, z)$ at time $t$, is proportional to $p(x_0, y_0, z_0, t_0)I(\theta)/[v_{rec}(t$ $- t_0)]^2$. In order to reach the point $P$ at time $t$, the position $A_0$ of NO molecules produced at time $t_0$ with $v_{rec}$ should be on a sphere $C_0$ of radius $v_{rec}(t - t_0)$ centered at $(0, 0, z - v_{mol}(t - t_0))$ where $v_{mol}$ is the molecular beam speed. Note that $v_{rec}$ is the recoil speed of NO in the frame moving at a speed of $v_{mol}$. Therefore, the number of NO molecules with a recoil speed $v_{rec}$ arriving at the probing position $P$ at time $t$ is given by

$$N_{tot}(z, t; v_{rec}) = g \int_{-\infty}^{t} \int_{C_o} \frac{p(x_0, y_0, z_0, t_0) I(\theta)}{[v_{rec}(t - t_0)]^2} dS_0 dt_0, \qquad (2)$$

with $g$ being a proportionality constant.

For the calculation of the spatial profile of the fragmented NO as a function of $z$ at a delay time $t$ using Eq. (2), the values of $v_{rec}$ and $\beta$ need to be determined. Fig. 2a shows an ion image of NO molecules ($^2\Pi_{1/2}$, $v = 1$, $J \sim 8.5$) produced by the photodissociation of $NO_2$. The polarizations of both lasers are adjusted to be perpendicular to the molecular beam direction. The angle $\theta$, therefore, is measured with respect to the y-axis. The time delay between the two laser pulses is set to 20 ns to avoid two-color multi-photon effects. This image is a 2-dimensional projection of the 3-dimensional ion velocity distribution obtained by velocity mapping under a focusing configuration of the ion lens. Hence the velocity distribution of NO molecules on the xy plane can be obtained from the image of the $NO^+$ spatial distribution using the magnification factor of the ion lens and time-of-flight of $NO^+$. Note that the average speed of ions should be the same as those of NO molecules since the ion recoil velocity vector due to the excess energy during the REMPI process are randomly oriented. To obtain $v_{rec}$ and $\beta$ values of fragmented neutral NO molecules, therefore, the raw image is transformed to a 2-dimensional cut image in Fig. 2b. The BASEX method is used for this inverse Abel transformation [23] to yield a recoil velocity $v_{rec}$ = 530 m/s with $\Delta v_{rec}$ ~ 100 m/s and a recoil anisotropy factor $\beta = 1.5$.

This recoil velocity is in excellent agreement with the theoretical recoil velocity given by $v_{rec} = \sqrt{2 m_O E_{excess} / m_{NO}(m_{NO} + m_O)}$, with $m_{NO}$ and $m_O$ the mass of NO and O, respectively. For the photofragmentation process $NO_2 + h\nu(355 \text{ nm}) \rightarrow NO(X^2\Pi_{1/2}, J \sim 8.5) + O(^3P_2)$, the excess energy is given by $E_{excess} = h\nu + E_{int} - D_0 - E_{rot}^{NO} - E_{vib}^{NO} - E_e^{NO} - E_e^{O}$, where $E_{int}$ is the internal energy of $NO_2$, $D_0$ the bond dissociation energy of $NO_2$, $E_{rot}^{NO}$ the rotational energy of NO, $E_{vib}^{NO}$ the vibrational energy of NO, $E_e^{NO}$ the electronic energy of NO, and $E_e^O$ the electronic energy of O. Inserting the known spectroscopic constants [24], a recoil velocity of 529 m/s is obtained. Several factors contribute to the recoil velocity spread of ~100 m/s. The initial transversal velocity, velocity changes during the REMPI process, and detector blurring cause velocity spread of about 70 m/s, which is estimated from the ion image of residual NO in equilibrium with $NO_2$ in the sample mixture. Additional contribution to the spread is from the photofragmentation process into $O(^3P_1)$, which has one fifth probability of the process into $O(^3P_2)$ [19, 25], producing an additional lobe in Fig. 2a corresponding to the recoil velocity of about 510 m/s. This unresolved lobe of smaller intensity increases the velocity spread by a few m/s. Based on the stagnation condition similar to ours [15], the rotational temperature can be assumed to be less than 10 K, which corresponds to ~14 cm$^{-1}$. This rotational energy is an order of magnitude smaller than the electronic energy difference 158 cm$^{-1}$ between $^3P_2$ and $^3P_1$ states and the effect of the initial rotational distribution of $NO_2$ contributes a few m/s to the spread.

The recoil anisotropy factor $\beta = 1.5$ is different from the previously reported values, 1.25 [22] and 1.2 ± 0.3 [19]. The anisotropy factor of NO depends on the rotational temperature of the $NO_2$ and there are a few potential problems in measuring it such as the Doppler shift and rotational alignment of the photofragments with respect to the relative velocity vector; the rotational temperature less than 10 K gives an anisotropy factor between 1.4 (estimated value with rotational temperature of 15 K) and 1.5 (asymptotic maximum at 0 K) [19]. The maximum Doppler shifts for NO molecules fragmented into the laser propagation axis are ±0.04 cm$^{-1}$. With the given dye laser bandwidth of 0.08 cm$^{-1}$ these NO molecules absorb only 55% of the ionization laser peak intensity, which results in overestimation of $\beta$ by about 0.2. Although there are some uncertainties due to the other factors such as the correlation between the recoil velocity vector of dissociated NO molecules and their rotational vector, the measured anisotropic factor value 1.5 is in agreement with the expected value from the initial rotational temperature of $NO_2$ and sufficient for our simulation using Eq. (2).

Next by making the polarization direction of the photodissociation UV laser parallel to the molecular beam axis ($z$ axis), the NO photofragments from $NO_2$ can be decelerated or accelerated. The polarization of the ionization dye laser was also adjusted to be parallel to the $z$ axis to make the laser polarization configuration the same as in Fig. 2 determining the $v_{rec}$ and $\beta$ values. Changing the relative $z$ position of the UV laser focus by ±300 $\mu$m with respect to the dye laser focus, the spatial profiles of the accelerated or decelerated NO molecules for a given delay time are obtained. At each probing position, 1500-laser-shot images of $NO^+$ ions from the two-color process of dissociation of $NO_2$ at 355 nm and then ionization of NO at 473 nm are obtained. Since the main source of the background signal is the one-color photofragmentation and ionization processes at 473 nm, the image of the two color process is corrected by subtracting the ion image obtained only with the 473 nm laser pulse. Then, the corrected signal imaged is binned 2-dimensionally and normalized with the ionization laser power at 473 nm.

Fig. 3 shows the temporal changes in spatial profiles of NO photofragments up to 120 ns after the peak of the photodissociation laser. As the delay time gets longer, the decelerated molecules and the accelerated molecules are separated more. Since the spatial and temporal overlap of the two photodissociation and ionization lasers causes 2-color multiphoton processes, 0-ns time delay experiment is omitted.

The NO profiles in solid lines in Fig. 3 are the simulation results obtained from Eq. (2) with the recoil anisotropy factor $\beta$ = 1.5, the recoil velocity $v_{rec}$ = 530 m/s, and the molecular beam velocity $v_{mol}$ = 565 m/s. The molecular beam velocity is estimated from the theoretical equation, $v_{mol} = (2c_p T_0/M)^{1/2}$, where $M$ is the number-averaged molecular weight of the gas mixture, $c_p$ is the number-averaged molar heat capacity of the gas mixture at constant pressure, and $T_0$ = 293 K is the initial temperature [26]. A dashed profile for a zero delay time is also depicted for comparison. Note that since at $t$ = 0 only a half of the dissociation laser pulse produces NO molecules, the area of this profile is smaller than the profiles at $t$ = 20 and 60 ns when a full dissociation laser pulse has produced NO molecules. All the simulation results are multiplied by a scaling factor to match the peak heights of the experimental and simulation profiles for $t$ = 20 ns. Assuming a delta-function-like initial velocity distribution for the molecular beam the most probable velocity along the $z$ axis ($v_{z,mp}$) of the decelerated molecules, numerically calculated using Eq. (1), is 144 m/s which agrees well with the slowly moving decelerated part of the spatial profiles in Fig. 3, in which the decelerated peak moves at about 130 m/s along the *z* axis. The accelerated parts of the profiles, which are moving at 980 m/s, also show a good agreement with the most probable accelerated velocity, 986 m/s, obtained from simulation. The profile area should decrease with the time delay since the fragmented molecules also spread along the *x*- and *y*- axes, as observed by the experimental data. Considering uncertainties in determining parameters, such as the molecular beam velocity, the recoil velocity, the waist radius, and the pulse width of the UV laser, and the fact that the numerical calculation neglects the initial velocity spread of the molecular beam and the finite size of the dye laser focus, the experimental and the numerical analysis results agree quite well.

The NO molecules photodissociated against the molecular beam direction ($\theta = \pi$) have a minimum longitudinal velocity $v_z^0 = 35$ m/s (= $v_{mol} - v_{rec}$). The NO molecules photodissociated at an angle $\theta = \pi - \delta$ have a longitudinal velocity $v_z^0 + \Delta v_z$ given by

$$\cos\delta = \frac{v_{rec} - \Delta v_z}{v_{rec}}. \tag{3}$$

The fraction of the decelerated molecules (*F*) with longitudinal velocity $v_z^0$ to $v_z^0 + \Delta v_z$ can be estimated by using the angular distribution given in Eq. (1):

$$F = \frac{\int_{\pi-\delta}^{\pi} I(\theta)\sin\theta d\theta}{\int_0^{\pi} I(\theta)\sin\theta d\theta}. \tag{4}$$

For $\Delta v_z = 10$ m/s ($\delta = 0.2$), Eq. (4) states that 2.9% of the NO molecules in the probed state are decelerated to have their longitudinal velocity in the range of 35-45 m/s. The deceleration fraction, evaluated from the ion image in Fig. 2a by dividing the image intensity integrated over $\pi - \delta < \theta < \pi$ with the total intensity, is 3.9%, which is slightly larger than the result from Eq. (4). This discrepancy is probably due to the contributions from the ionization of NO molecules in a number of rotational states. Since about 40% of the NO fragments are in the first vibrationally excited state [22], assuming that about 50 rotational states are equally populated, 0.03% of the NO molecules produced by photodissociation would be decelerated in this velocity range.

The decelerated NO molecules stay for a relatively long time at the origin defined by the dissociation laser focal center. The three images in Fig. 4b are obtained at 20 ns, 100 ns, and 300 ns after the photodissociation at the origin. The NO molecules with a large transverse velocity leave the origin quickly, which makes it possible to separate the NO molecules produced with the recoil velocity parallel to the $z$ axis from the others. The experimental data in Fig. 4a are obtained by binning images, such as the ones in Fig. 4b, at given times. It shows the relative changes in their numbers at the origin. The experimental data (circles) and the numerical simulation results using Eq. (2) (solid line) are in excellent agreement. The spreads (FWHM) of the transverse velocities are ±320, ±190, and ±130 m/s for 20, 100, and 300 ns delays, respectively, which correspond to $\delta$ values of 0.6, 0.4, and 0.2. Using Eq. (3), the $\Delta v_z$ values of NO molecules 100 and 300 ns after production are estimated to be 32 and 14 m/s, respectively. The $\Delta v_z$ value for a 20 ns delay has not been estimated since both the accelerated and decelerated components are involved as shown in Fig. 3. The decelerated and propagating along the $z$ axis might be manipulated with a molecule prism [16]. When the molecular beam velocity is reduced from 565 m/s to 35 m/s, the velocity change due to the molecule prism of a laser pulse with a 5 $\mu$m waist radius and 10 ns pulse width will be enhanced by 80%.

4. Conclusions

We have demonstrated that photodissociation can be exploited to produce slow ($v_z^0 \sim$ 35 m/s) neutral molecules by partially canceling the molecular beam velocity 565 m/s. The spatial profiles of the NO molecules from the photodissociation of $NO_2$, showing two peaks the decelerated one having $v_{z,mp} \sim$ 130 m/s and the accelerated one having $v_{z,mp} \sim$ 980 m/s, agree well with the simulation results. Therefore, the NO molecule specifically dissociated into $-z$ direction has average velocity of ~35 m/s. Although the achieved minimum velocity 35 m/s is not small enough to give the quantum effect mentioned in the introduction, in principle this simple method can produce much slower molecules when the initial molecular beam velocity is properly controlled through the choice of carrier gas and temperature. The decelerated NO molecules stayed around the origin even 300 ns after the production by photodissociation of $NO_2$.

**Acknowledgments**

This work was supported by the Korea Research Foundation Grant (KRF-2006-311-C00263) and the Seoul R&BD Program.

**Figure captions**

FIG. 1. (a) Schematics of the setup. (b) NO molecules produced at $A_0(x_0, y_0, z_0)$ and at time $t_0$ with a recoil velocity $v_{rec}$ lie on a sphere centered at $A(x_0, y_0, z_0 + v_{mol}\Delta t)$ and at time $t$ (= $t_0$ + $\Delta t$). For these NO molecules to reach the probing point $P(0, 0, z)$ at time $t$, $A_0$ should be on the sphere $C_0$ (dotted line) centered at $P_0(0, 0, z - v_{mol}\Delta t)$. The center of mass of the system (NO + O) is on the sphere $C$ (gray solid line) at time $t$. The origin $O$ is the focal center of the UV laser (gray dotted line). The angle $\theta$ depicts the angle $\theta$ in Eq. (1) for the case in which the dissociation laser polarization is parallel to the $z$-axis.

FIG. 2. (a) Ion image of the photodissociated NO($^2\Pi_{1/2}$, $v = 1$, $J \sim 8.5$) and (b) its cross section reconstructed by an inverse Abel transformation. Polarizations of the dissociation and the ionization lasers are parallel to the $y$-axis and the time delay between them is 20 ns. The size of the images is 480 × 480 pixels, which corresponds to 1750 × 1750 m/s.

FIG. 3. Spatial profiles of the photodissociated NO at (a) 20, (b) 60, and (c) 120 ns after dissociation, respectively. The power of the UV laser and dye laser are 0.4 and 5 mW, respectively. Circles are experimental data and solid lines are simulation results. For comparison, a theoretical profile with zero delay between the two lasers is shown with a dashed line (see text).

FIG. 4. (a) Temporal profile of the relative number density of the NO at the origin and (b) three images at 20, 100, and 300 ns after the photodissociation of $NO_2$. The linear polarizations of the lasers are fixed along the $z$ axis perpendicular to the figure plane.

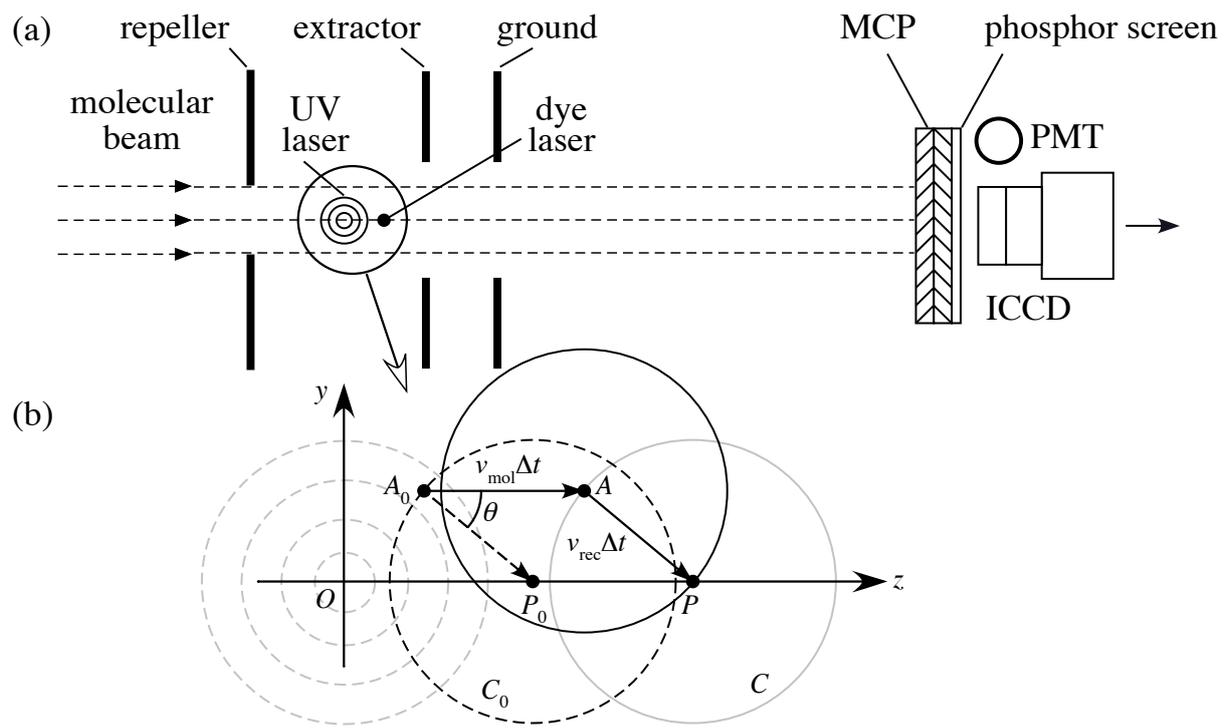

Figure 1

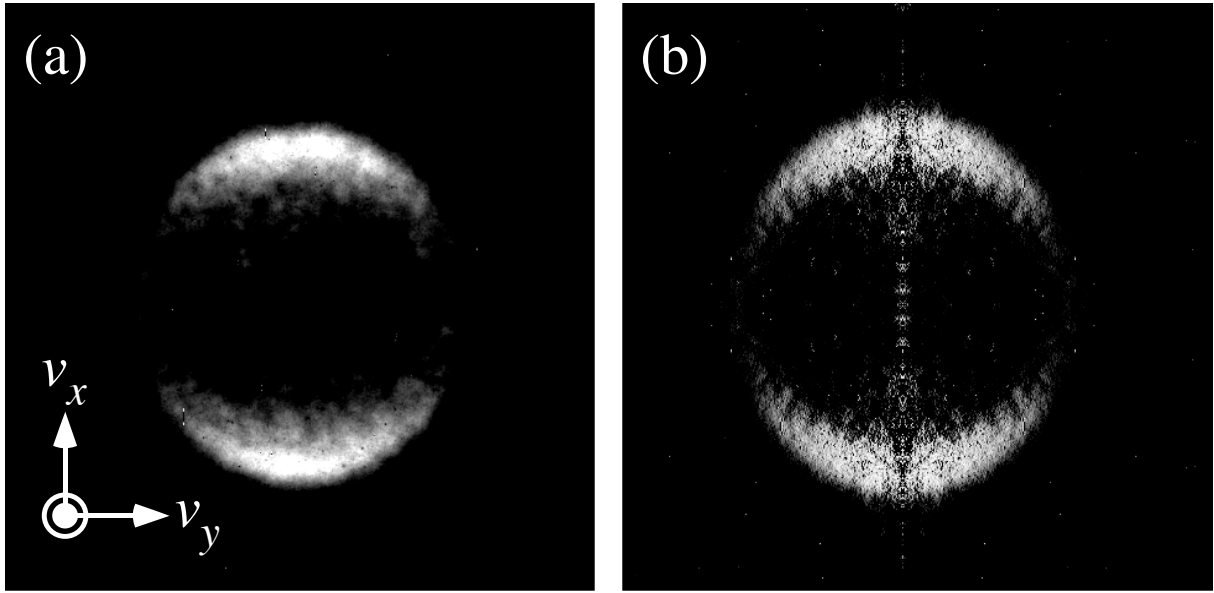

Figure 2

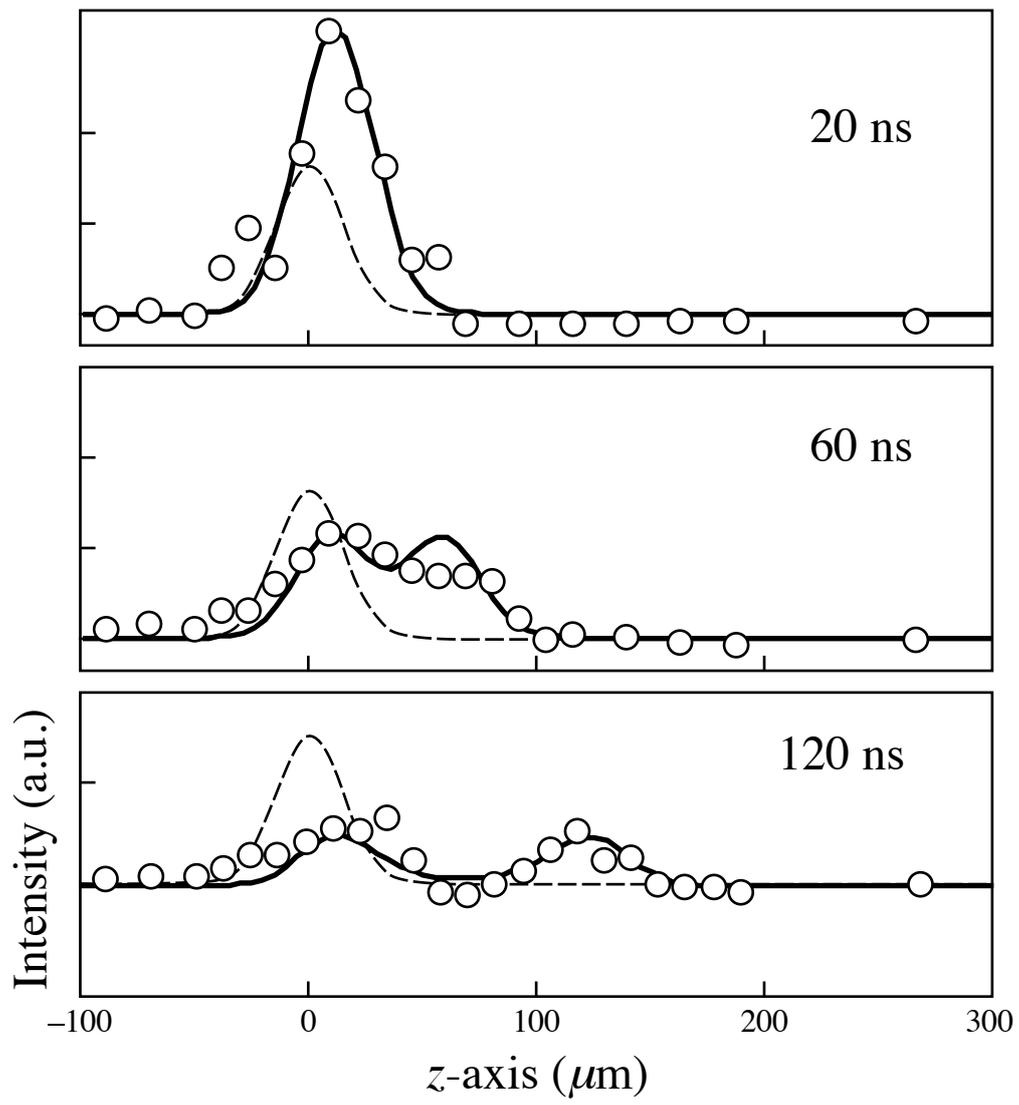

Figure 3

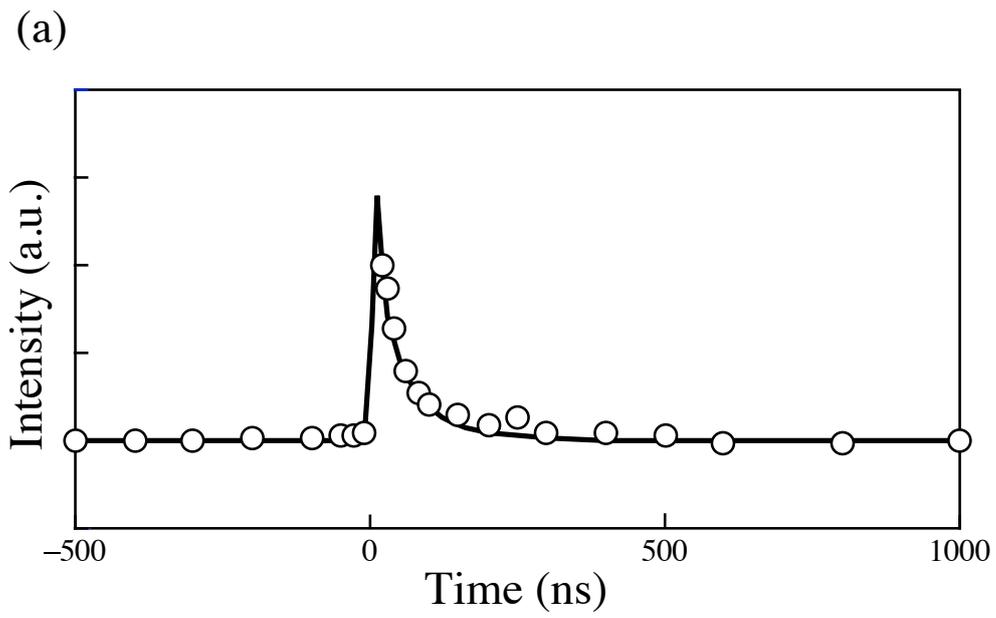

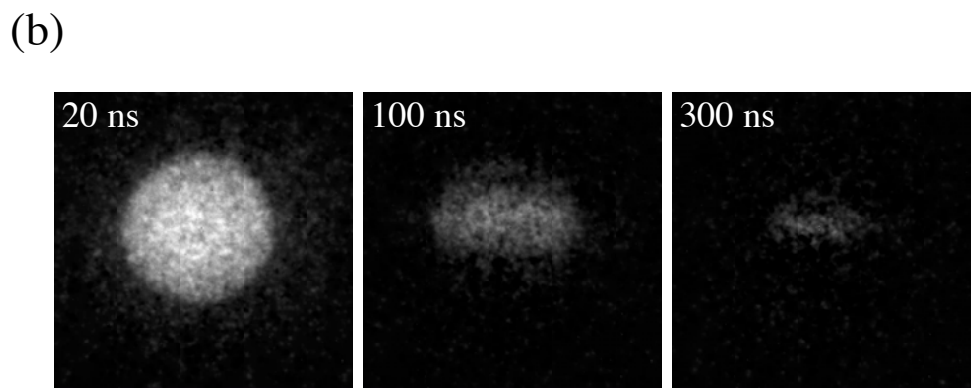

Figure 4